# ESSnuSB status

George Fanourakis
Institute of Nuclear & Particle Physics, NCSR Demokritos,
Agia Paraskevi, Attiki, Greece
On behalf of the ESSnuSB/ESSnuSB+ collaboration
Contribution to the 25th International Workshop on Neutrinos from Accelerators

**Abstract**

ESSnuSB (the European Spallation neutrino Super Beam) is a design study for a long-baseline neutrino experiment to precisely measure the CP violation in the leptonic sector, at the second neutrino oscillation maximum, using a beam driven by the uniquely powerful ESS proton linear accelerator. The ESSnuSB Conceptual Design Report showed that after 10 years, about 72% of the possible CP-violating phase, $δ_{CP}$, range will be covered with 5σ C.L. to reject the no-CP-violation hypothesis. The expected precision for $δ_{CP}$ is better than 8° for all $δ_{CP}$ values, making it the most precise proposed experiment in the field. The extension project, the ESSnuSB+, aims in designing two new facilities, a Low Energy nuSTORM and a Low Energy Monitored Neutrino Beam to be used to precisely measure the neutrino-nucleus cross-section in the energy range of 0.2−0.6 GeV. A new water Cherenkov detector will also be designed to measure cross sections and serve to explore the sterile neutrino case. An overall status of the project is presented together with the ESSnuSB+ additions.

**Introduction**

The next generation long base line neutrino experiments [1], currently in preparation, are designed to resolve the question of neutrino mass ordering, to determine the quadrant of the atmospheric mixing angle $θ_{23}$ via precision measurements and to prove the existence of CP violation (CPV). However, their precision in measuring the CP violating phase angle is not sufficient to resolve the theoretical models proposed to explain the lack of antimatter in the Universe [2,3]. ESSnuSB is a next-to-next generation experiment under design, with the main purpose to measure the CPV phase angle to such high precision as to enable the selection of the correct theory.

The probability of neutrino oscillations in vacuum is given by [4]:

$$P(\nu_\alpha \to \nu_\beta) = \delta_{\alpha\beta} - 4 \sum_{I<j} Re(U_{\alpha i}U_{\beta i}^* U_{\alpha j}^* U_{\beta j}) X_{ij} \pm 2 \sum_{I<j} Im(U_{\alpha i}U_{\beta i}^* U_{\alpha j}^* U_{\beta j}) X'_{ij} \qquad (1)$$

where $X_{ij} = sin^2(\Delta m_{ij}^2 L/4E)$, $X'_{ij} = sin(\Delta m_{ij}^2 L/2E)$, $U_{\alpha j}$ are the elements of the PMNS [5] matrix, the + sign is for neutrino oscillations, while the − sign is for antineutrino oscillations, E the energy of the neutrino or antineutrino, L the distance between the near and far detectors and $\Delta m_{ij}^2 = m_i^2 - m_j^2$ the splitting between the square masses of $\nu_i$ and $\nu_j$.

The determination of the CP violating phase angle relies on the measurement of the neutrino-antineutrino (matter-antimatter) asymmetry A, which in the case of ESSnuSB design is:

$$A = \frac{P(\nu_\mu \to \nu_e) - P(\bar\nu_\mu \to \bar\nu_e)}{P(\nu_\mu \to \nu_e) + P(\bar\nu_\mu \to \bar\nu_e)} \qquad (2)$$

The equation (1) is modified by matter effects, in the case of neutrinos traveling through matter (the case in all LBL neutrino experiments). The matter effects are not important for ESSnuSB because the neutrino energy spectrum is partly on the second oscillation maximum (see next section).

**The ESSnuSB project**

ESSnuSB plans to exploit the European Spallation Source (ESS) proton linear accelerator (linac), designed for 2 GeV proton kinetic energy and 5MW power, to create a super intense neutrino beam, with an energy distribution from 200 MeV/v to 600 MeV/c, and design a long Baseline (LBL) experiment tuned to measure the CPV angle to an unprecedented precision. The Conceptual Design Report (CDR) of this study has been completed and published in 2022 [6]. The study intends to establish a parallel experimental facility at ESS, without interfering with ESS's main mission to become the most powerful neutron facility in the world. The proton linac power is proposed to be doubled so to have equal power for the neutron and neutrino programs. The neutrino program requires the proton pulses to be shortened from 2.86 ms to 1.2 µs. This is accomplished by an accumulator which also requires that the linac accelerates H$^-$ instead of protons. A target station is designed with four targets and four horns to ease the effects of the high-power proton beam. A long baseline setup is proposed for the neutrino oscillation studies: A far water Cherenkov detector consisting of two tanks of 270 ktons each, placed in the Zinkgruvan mine, 360 km away, at a depth of about 1 Km, and a near detector complex consisting of a water Cherenkov, a super fine-grained scintillator detector and an emulsion detector (see Fig. 1).

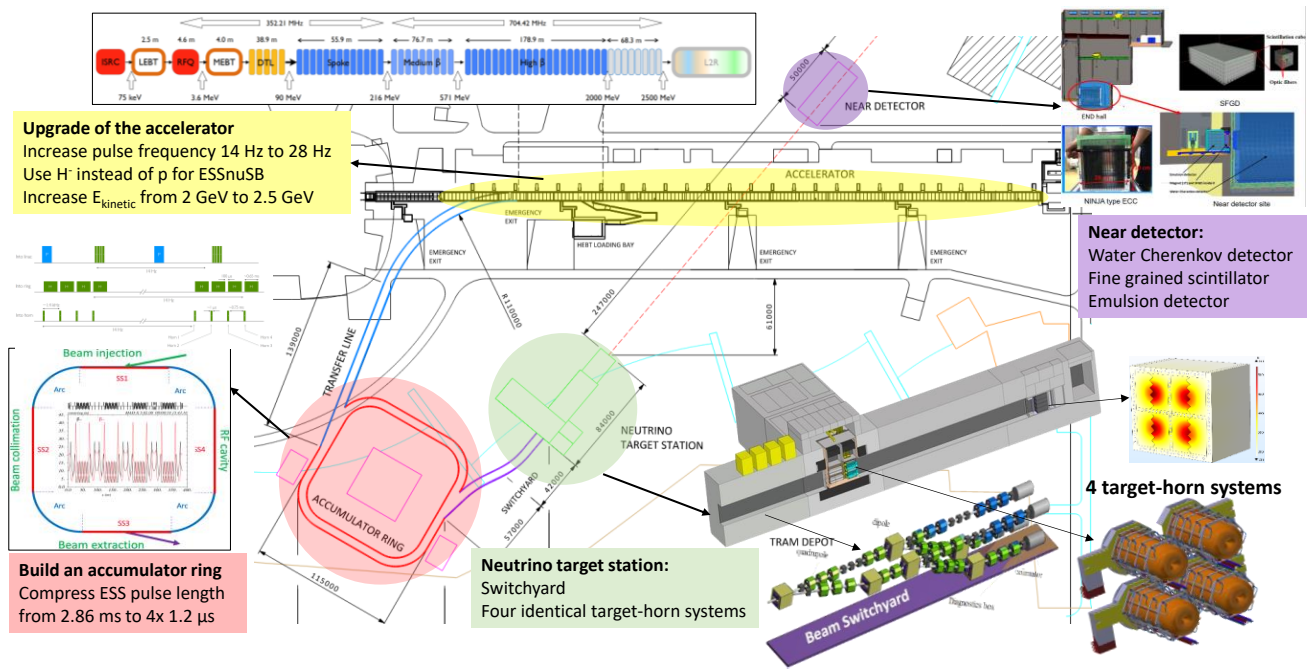

**Figure 1.** Layout of the proposed modifications and additions to ESS beam line and Instrumentation.

The Physics reach of ESSnuSB covers the typical program of the LBL neutrino experiments, measuring the neutrino oscillation parameters, especially focusing on the precision measurement of the CPV phase. Its ability to obtain a high precision is because a large part of the ESSnuSB neutrino spectrum is sitting at the second oscillation maximum a fact that is providing the following benefits: a) It makes the matter-antimatter asymmetry A$_{CPV}$ about 2.5 times larger than for the case of spectra at the first maximum [7], due to a larger interference term, a fact which also makes the measurement of the CPV more stable against systematical errors, and b) the matter effects of neutrino interactions on their way to the far detector are much smaller at the second maximum (Fig. 2) without danger of faking CPV as when working in the first maximum [8].

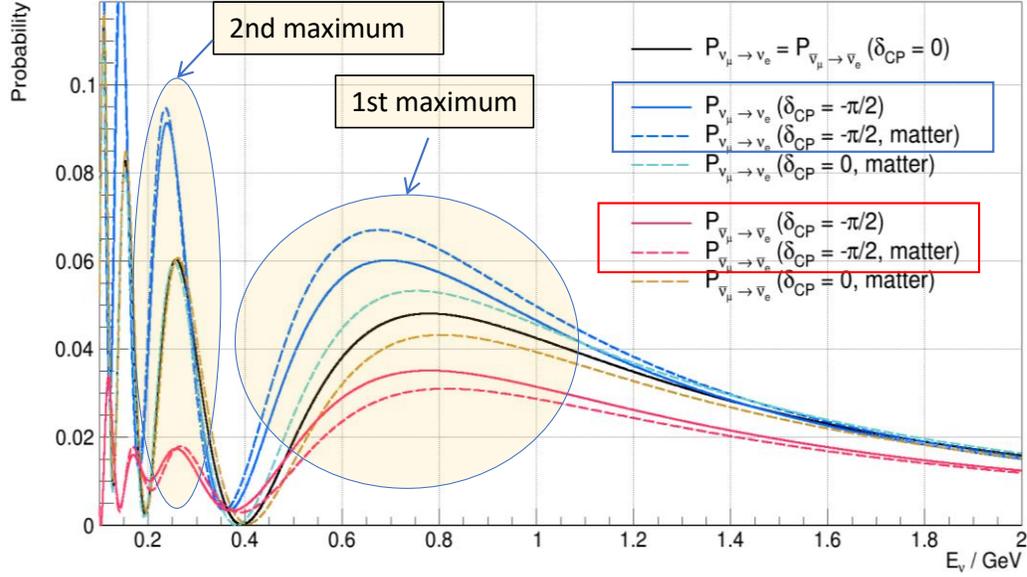

**Figure 2.** Neutrino and antineutrino disappearance oscillation probabilities as a function of neutrino energy at the fixed distance of 360 km. The oscillation probabilities are shown for $\delta_{CP} = 0$ and $\delta_{CP} = \pi/2$. Full lines correspond to oscillations in vacuum and dashed lines to oscillations in matter.

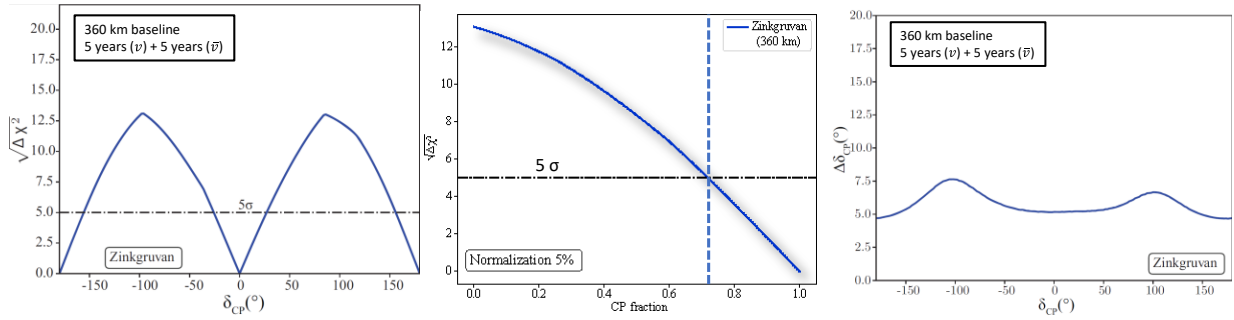

**Figure 3.** (Left panel) Discovery sensitivity of the leptonic CPV as a function of $\delta_{CP}$ range. (Middle) The fraction of the covered $\delta_{CP}$ range for which CPV can be discovered as a function of the significance. (Right) The precision on $\delta_{CP}$ value as a function of the $\delta_{CP}$ range.

The main physics results obtained from this study are shown in Fig. 3. The results assume 5% normalization uncertainty, 5 years run with neutrino beam and 5 years with antineutrinos. The left plot shows the experiment's sensitivity for CP-violation discovery as a function of $\delta_{CP}$ range of true values. It shows that at the maximal violation, i.e., $\delta_{CP} \sim \pm 90°$, the CPV discovery sensitivity reaches about $12\sigma$. The middle plot shows that ESSnuSB will cover more than 70% of the range of the true $\delta_{CP}$ values with a confidence level of more than $5\sigma$ to reject the no-CPV hypothesis, after 10 years of data taking. The right plot shows that the expected precision on the measured value of $\delta_{CP}$ will be better than 8° for the whole $\delta_{CP}$ range.

**The ESSnuSB+ project**

After the successful completion of the ESSnuSB design study, we have proposed the ESSnuSB+ project [9]. Its purpose is to extend the Instrumentation arsenal of ESSnuSB, by adding two more neutrino beam devices and a near-near detector for precise low energy neutrino cross section measurements, as well as to expand and improve the Physics potential of ESSnuSB. The precise knowledge of the neutrino cross sections for the ESSnuSB neutrino beam spectrum is necessary to reduce the systematics of the measurements of the neutrino oscillation parameters. The new project is funded by the European commission Horizon-Europe program for the period 2023−2026.

The design of a new target station is foreseen for the new project, to generate pions which will feed a Low Energy nuSTORM (LEnuSTORM), i.e., a muon racetrack storage ring (see Fig. 4-left panel), like the nuSTORM facility planned for CERN [10], and a Low Energy Monitored neutrino beam (LEMNB) decay tunnel (see Fig. 4-right panel) inspired by the ENUBET project [11]. Both facilities will constitute the first phase of staging ESSnuSB, the main long baseline experiment, a phase which will involve the precision neutrino cross section measurements with a new under design near-near detector, the LEMMOND water Cherenkov detector (Low Energy neutrino from stored Muons and Monitored beam Near Detector). The decay of negative polarity muons in the straight sections of LEnuSTORM will provide clean equal amounts of muon neutrinos and electron antineutrinos or muon antineutrinos and electron neutrinos when positive muons are used. LEMNB will also provide clean neutrino and antineutrino beams by tagging the muons from pion decays. These well-defined neutrino beams will be used with the LEMMOND detector, to precisely measure low energy neutrino cross sections with water.

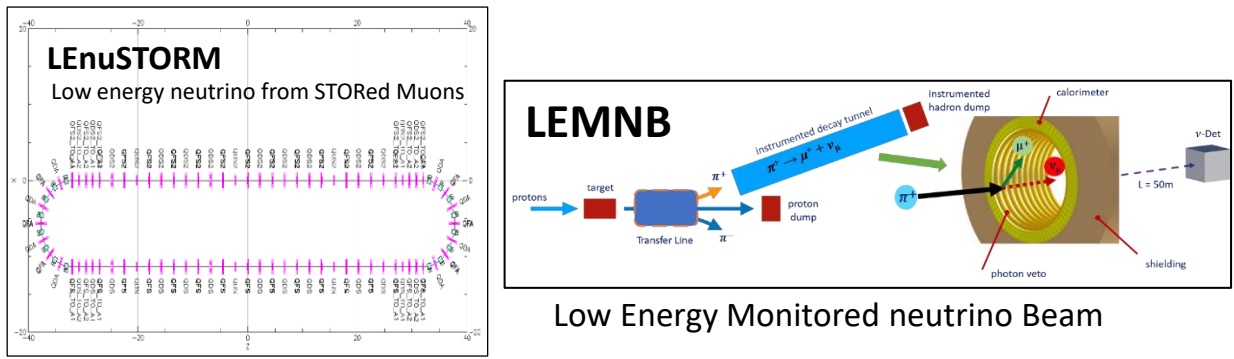

**Figure 4.** (Left) The low energy neutrino beam from stored muons in a racetrack storage ring.
(Right) The instrumented pion decay tunnel for the low energy monitored neutrino beam (LEMNB).

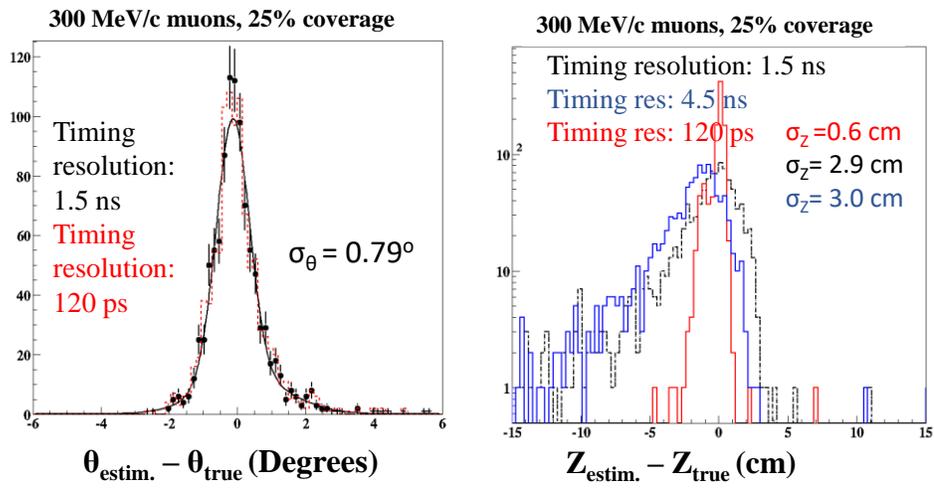

**Figure 5.** A less than 1° θ resolution (left) was obtained for a toy detector consisted of a single sensor plane and a <3 cm (<1 cm) resolution for the z position of the muon verted was obtained (right) for 1.5 (120 ps) sensor time resolutions.

LEMMOND, the new detector added to the ESSnuSB detector suite, is a cylindrical Cherenkov with about 5 m diameter and 10 m length. It will be situated 50 m away from the new neutrino facilities LEnuSTORM or LEMNB. An initial design of the simulation and reconstruction tools was tested for a simplified geometry of a flat detector plane made of 6400 5x5 cm$^2$ sensors. GEANT4 simulated muon tracks were generated with angles θ=0º, 30º and ϕ=0º, 200 cm away from the detector plane and their Cherenkov photons emission was followed to the sensors. A quantum efficiency of 25% and 25% coverage was assumed. Two choices for the time resolutions of the sensor elements were considered, 1.5 ns and 120 ps.

The results of the analysis are quite encouraging, as Fig. 5 shows. Theta (θ) and phi (ϕ) angle resolutions of less than 1º were obtained for both timing capability choices. For the determination of the interaction point (the Z coordinate of the muon vertex) a precision of about 6 cm was obtained for the 1.5 ns sensors and less than 1 cm for the 120 ps sensors. These preliminary encouraging results will be hopefully closely reproduced with full cylindrical geometry and the neutrino interactions simulation and reconstruction will also be developed with the ongoing work.

The LEnuSTORM and the combination of the LEMMOND detector with the near detector complex of ESSnuSB constitute an excellent Short Base Line (SBL) setup able to investigate the existence of sterile neutrinos of mass square differences of 1 eV$^2$ to 10 eV$^2$.

Besides the use of the new infrastructure for neutrino cross section measurements and the new option added of a SBL experiment for sterile neutrino investigations, further analysis techniques and physics cases are being developed within the new project. A Monte Carlo study of neutrino oscillations using atmospheric neutrinos in the far detector [12] showed that in 4 years ESSnuSB could determine the correct mass ordering with 3σ significance, could determine the $θ_{23}$ octant at 3σ in 4(7) years for normal (inverted) ordering and provide constraints on $θ_{23}$ and $Δm^2_{31}$ (see Fig. 6). The addition of Gadolinium in the Cherenkov detectors and the enhanced capability of final state neutron detection is being investigated.

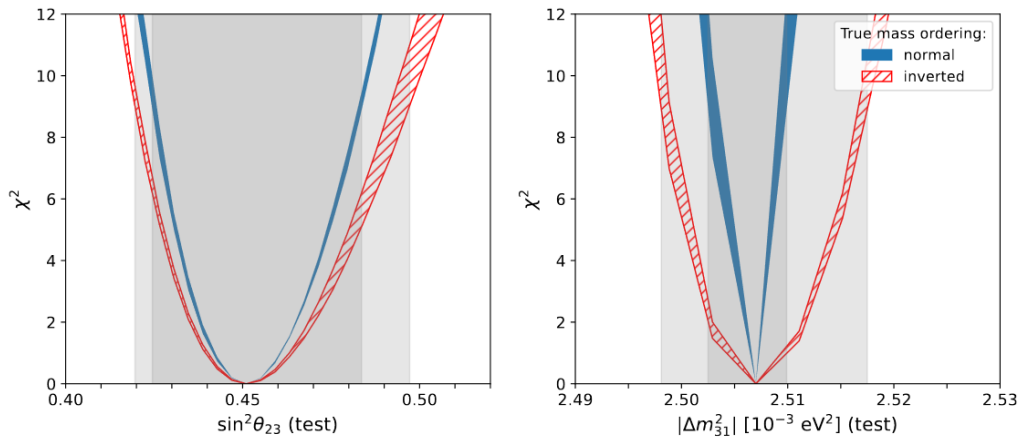

**Figure 6.** Constraints on $θ_{23}$ (left) and $Δm^2_{31}$ (right). Shaded areas indicate the allowed values for normal (dark) and inverted (light) ordering.

The sensitivity of ESSnuSB for Beyond the Standard Model (BSM) is also being investigated within the new project. Results have been published for constraints on scalar Non-Standard Interactions parameters [13] and for constraints on Quantum Decoherence parameters [14].

**Conclusions**
ESSnuSB and ESSnuSB-plus are the two phases of a next-to-next generation neutrino oscillation facility proposed which includes a) the precision measurement of the CP Violating phase with a long base-line neutrino setup including a near detector suite, made of a water Cherenkov, a Super fine grained detector and an emulsion detector, and an underground far water Cherenkov detector, b) the design and construction of novel facilities, including a LEnuSTORM ring, a LENMB instrumented neutrino tunnel and a near-near water Cherenkov detector, to precisely measure low energy neutrino cross sections and c) investigate the existence of sterile neutrinos with a short base-line setup. More Physics capabilities are being evaluated during the second phase of the ESSnuSB project.

**Acknowledgements**

This work is funded by the European Union. Views and opinions expressed are however those of the author(s) only and do not necessarily reflect those of the European Union. Neither the European Union nor the granting authority can be held responsible for them. We acknowledge further support provided by the following research funding agencies: Centre National de la Recherche Scientifique, France; Deutsche Forschungsgemeinschaft, Projektnummer 423761110 and the Excellence Strategy of the Federal Government and the Länder, Germany; Ministry of Science and Education of Republic of Croatia grant No. KK.01.1.1.01.0001; the European Union's Horizon 2020 research and innovation programme under the Marie Skłodowska -Curie grant agreement No 860881-HIDDeN; the European Union NextGenerationEU, through the National Recovery and Resilience Plan of the Republic of Bulgaria, project No. BG-RRP-2.004-0008-C01.